\documentstyle[array,aps,epsf]{revtex}
\begin{document}
\draft
\title
{Relativistic entanglement and Bell's inequality}
\author{Doyeol Ahn$^{1,2}${\footnote{e-mail:dahn@uoscc.uos.ac.kr}},
Hyuk-jae Lee$^1${\footnote{e-mail:lhjae@iquips.uos.ac.kr}}, Young Hoon Moon$^{1,2}$and Sung
Woo Hwang$^{1,3}${\footnote{e-mail:swhwang@korea.ac.kr}}}
\address{
$^1$Institute of Quantum Information Processing and Systems,
University of Seoul, Seoul, 130-743, Korea\\
$^2$Department of Electrical and computer Engineering, University
of Seoul,
Seoul, 130-743, Korea\\
$^3$Department of Electronic Engineering, Korea University, Seoul,
136-701, Korea}

\maketitle

\vspace{2.0cm}



\begin{abstract}
In this paper, the Lorentz transformation of entangled Bell states seen by a moving observer is studied. The calculated Bell observable for
$4$ joint measurements turns out to give a universal value, $\langle\hat{a}\otimes\vec{b}\rangle+\langle\hat{a}\otimes\vec{b}'\rangle+\langle\hat{a}'\otimes\vec{b}\rangle
-\langle\hat{a}'\otimes\vec{b}'\rangle=\frac{2}{\sqrt{2-\beta^2}}(1+\sqrt{1-\beta^2})$, where $\hat{a}, \hat{b}$ are the relativistic spin observables
derived from the Pauli-Lubanski pseudo vector and $\beta=\frac{v}{c}$.

We found that the degree of violation of the Bell's inequality is decreasing with increasing velocity of the observer and the Bell's inequality
is satisfied in the ultra-relativistic limit where the boost speed reaches the speed of light.
\end{abstract}
\vspace{.25in}


\newpage
\section{Introduction}
Relativistic transformation properties of quantum states, especially, the entangled states, are of considerable
interest lately partially because many novel features of the quantum information processing rely on the entanglement
and the nonlocality associated with it \cite{czac}--\cite{hwan}. One would take the teleportation \cite{benn},\cite{bouw} as a typical example. The problem of the
entanglement and the nonlocality traces back to the famous 1935 paper by Einstein-Podolsky-Rosen\cite{eins}, almost seventy years
ago, now known as the EPR paradox, and subsequent studies, most noticeably by Bell\cite{bell}, showed that the Nature indeed
seems to be nonlocal as far as non-relativistic quantum mechanics is concerned. This subtle question still
remains to be answered especially in the relativistic arena. Recently, Czarchor\cite{czac} and Terashimo and Ueda\cite{tera} suggested
that the degree of violation of the Bell inequality depends on the velocity of the pair of spin-$\frac{1}{2}$ particles or
the observer with respect to the laboratory.

The goal of this article is to try to give a partial answer to the EPR paradox and the nonlocality. In the previous work\cite{lee},
we studied the case of an entangled state shared by Alice and Bob in different frames and showed that the entangled pair
satisfies the Bell's inequality when the boost speed approaches the speed of light somewhat surprisingly. We also
showed that the Bell state in the rest frame appears as a superposition of the Bell bases to an observer in the moving
frame due to the Wigner rotation of the spin states.

In this paper, we calculate the Bell observables for entangled states in the rest frame seen by the observer moving in the
$x$-direction and show that the entangled states satisfy the Bell's inequality when the boost speed approaches the speed
of light. The calculated average of the Bell observable for the Lorentz transformed entangled states turns out to be
\begin{eqnarray}
c(\vec{a},\vec{a}',\vec{b},\vec{b}')&=&\langle\hat{a}\otimes\hat{b}\rangle+\langle\hat{a}\otimes\hat{b}'\rangle+\langle\hat{a}'\otimes\hat{b}\rangle
-\langle\hat{a}'\otimes\hat{b}'\rangle\nonumber\\
&=&\frac{2}{\sqrt{2-\beta^2}}(1+\sqrt{1-\beta^2}),
\label{bell}
\end{eqnarray}
where $\hat{a}, \hat{b}$ are the relativistic spin observables for Alice and Bob, respectively, related to the
Pauli-Lubanski pseudo vector \cite{czac},\cite{ryde} which is known to be a relativistically invariant operator corresponding to spin and
$\beta=\frac{v}{c}$, the ratio of the boost speed and the speed of light.

In the non-relativistic limit $\beta=0$, and the right half side(r.h.s.) of eq. (\ref{bell}) gives the value of
$2\sqrt{2}$, indicating the maximum violation of the Bell's inequality. On the other hand, in the ultra-relativistic limit,
$\beta=1$, the r.h.s. of (\ref{bell}) gives the value $2$, suggesting the Bell inequality is satisfied. Moreover,
the equation (\ref{bell}) shows that the average Bell observable or the degree of violation of the Bell's
inequality is decreasing with increasing velocity of the observer.

In the next section, we give the Lorentz transformation of the quantum states and the Wigner representation of the Lorentz
group from a heuristic point of view.

\section{Relativistic transformation of quantum states and Wigner representation of the Lorentz group}
One of the conceptual barriers for the relativistic treatment of quantum information processing is the difference of the
role played by the wave fields and the states vectors in relativistic quantum theory. In non-relativistic quantum mechanics
both the wave function and the states vector in Hilbert space give the probability amplitude which can be used to define
conserved positive densities or density matrix. Since an attempt
to unify quantum mechanics and special relativity was made by Dirac toward the end of the 1920's and the famous
Dirac equation for an electron was discovered, it was found that the waves obeying the relativistic wave equations
do not represent the probability amplitude by themselves. For example, the probability wave function for a photon is neither
the electric nor the magnetic field which satisfies the Maxwell's equations. In a way, the state vector of a photon
is related indirectly to the Maxwell's equations. In this sense, the relativistic wave equations must be regarded only as indirect
representation for the description of one-particle probability waves, and the forms of equation themselves have a direct
connection to the quantum field theory.

One the other hand, the idea that the quantum states of relativistic particles can be formulated directly without the use
of wave equations, was proposed by Wigner in 1939\cite{wigne}. He showed that the states of a free particle are given
by unitary irreducible representations of the Poincar\'{e} group, i.e. the group formed by translations
and Lorentz transformations in the Minkowski space. As a matter of fact, if we get all unitary irreducible representations of
the Poincar\'{e} group, or the Lorentz group, we do have a complete knowledge of relativistic free particle states and behavior \cite{ohnu}.

In this paper, we follow Wigner's approach and focus on the Lorentz transformation properties of quantum states,
then obtain the relativistic transformation of entangled quantum states. For convenience, we follow Weinberg's notation \cite{wein}
throughout the article.

A multi-particle state vector is denote by
\begin{equation}
\Psi_{p_1\sigma_1;p_2\sigma_2;\dots}=a^+(\vec{p}_1,
\sigma_1)a^+(\vec{p}_2, \sigma_2)\dots\Psi_0,
\label{states}
\end{equation}
where $p_i$ labels the four-momentum, $\sigma_i$ is the spin $z$ component, $a^+(\vec{p}_i, \sigma_i)$ is the creation operator
which adds a particle with momentum $\vec{p}_i$ and spin $\sigma_i$, and $\Psi_0$ is the Lorentz invariant vacuum state.
The Lorentz transformation $\Lambda$ induces unitary transformation on vectors in the Hilbert space
\begin{equation}
\Psi\rightarrow U(\Lambda)\Psi
\end{equation}
and the operators $U$ satisfies the composition rule
\begin{equation}
U(\bar{\Lambda})U(\Lambda)=U(\bar{\Lambda}\Lambda),
\end{equation}
while the creation operator has the following transformation rule
\begin{equation}
U(\Lambda)a^+(\vec{p},\sigma)U(\Lambda)^{-1}=\sqrt{\frac{(\Lambda p)^0}{p^0}}\sum_{\bar{\sigma}}
{\cal D}^{(j)}_{\bar{\sigma}\sigma}(W(\Lambda, p))a^+(\vec{p}_{\Lambda},\sigma).
\label{trans}
\end{equation}
Here, $W(\Lambda, p)$ is the Wigner's little group element given by
\begin{equation}
W(\Lambda, p)=L^{-1}(\Lambda p)\Lambda L(p),
\label{litt}
\end{equation}
with ${\cal D}^{(j)}(W)$ the representation of $W$ for spin $j$, $p^{\mu}=(\vec{p}, p^0)$, $(\Lambda p)^{\mu}=
(\vec{p}_{\Lambda}, (\Lambda p)^0)$ with $\mu=1, 2, 3, 0$ and $L(p)$ is the standard Lorentz transformation such that
\begin{equation}
p^{\mu}=L^{\mu}_{{}\nu}k^{\nu}
\end{equation}
where $k^{\nu}=(0, 0, 0, m)$ is the four-momentum taken in the particle's rest frame. One can also use the conventional
ket-notation to represent the quantum states as
\begin{eqnarray}
\Psi_{p, \sigma}&=&a^+(\vec{p}, \sigma)\Psi_0\nonumber\\
&=&|\vec{p}, \sigma\rangle\nonumber\\
&=&|\vec{p}\rangle\otimes|\sigma\rangle.
\end{eqnarray}
We now give the derivation of the representation of the Wigner's little group $W(\Lambda, p)$ for spin-$\frac{1}{2}$
particles following Halpern's approach \cite{halp}.
>From eq. (\ref{litt}), the representation ${\cal D}^{(\frac{1}{2})}(W(\Lambda, p))$ is written as
\begin{equation}
{\cal D}^{(1/2)}(W(\Lambda, p))={{\cal D}^{(1/2)}}^{-1}(L(\Lambda p)){\cal D}^{(1/2)}(\Lambda){\cal D}^{(1/2)}(L(p)).
\label{dwlp})
\end{equation}
If we consider an arbitrary boost given by the velocity $\vec{v}$ with $\hat{e}$ as the normal
vector in the boost direction, the Lorentz transformation $\Lambda^{\mu}_{{}\nu}$ is \cite{gold}
\begin{eqnarray}
\Lambda^i_{{}j}&=&\delta_{ij}+e_i e_j (\cosh \alpha -1),\nonumber\\
\Lambda^i_{{}0}&=&\Lambda^0_{{}i}=e_i  \sinh \alpha ,\nonumber\\
\Lambda^0_{{}0}&=&\cosh\alpha\nonumber\\
&=&\gamma=\frac{1}{\sqrt{1-\beta^2}}.
\label{lamb}
\end{eqnarray}
Then, for $p^{\mu}=(\vec{p}, p^0)$ with $p^0=E_{\vec{p}}$,
\begin{eqnarray}
\vec{p'}&=&\vec{p_{\Lambda}}\nonumber\\
&=&[\vec{p}-(\vec{p}\cdot\hat{e})\hat{e}]+[E_{\vec{p}}\sinh\alpha+
(\vec{p}\cdot\hat{e})\cosh\alpha]\hat{e},\label{lore}\\
(\Lambda p)^0&=& (p^0)'\nonumber\\
&=&E_{\vec{p}}\cosh\alpha+(\vec{p}\cdot\hat{e})\sinh\alpha,
\end{eqnarray}
and
\begin{equation}
{\cal D}^{(1/2)}(\Lambda)=I\cosh\frac{\alpha}{2}+
(\vec{\sigma}\cdot\hat{e})\sinh\frac{\alpha}{2}.
\label{dlam}
\end{equation}
>From the 2-component spinor representation (Appendix):
\begin{eqnarray}
\phi_R(\vec{p})&=&[(\frac{\gamma+1}{2})^{1/2}+\vec{\sigma}\cdot\frac{\vec{p}}{|\vec{p}|}(\frac{\gamma-1}{2})^{1/2}]\phi_R(0)\nonumber\\
&=&{\cal D}^{(1/2)}(L(p))\phi_R(0),
\label{spino}
\end{eqnarray}
where $\phi_R$ is the 2-component spinor, we obtain
\begin{equation}
{\cal D}^{(1/2)}(L(p))=(\frac{p^0 +m}{2m})^{1/2}I+
(\frac{p^0-m}{2m})^{1/2}\vec{\sigma}\cdot\frac{\vec{p}}{|\vec{p}|},
\label{dlp}
\end{equation}
and
\begin{equation}
{\cal D}^{(1/2)}(\Lambda L(p))=(\frac{(\Lambda p)^0
+m}{2m})^{1/2}I+(\frac{(\Lambda p)^0-m}{2m})^{1/2}
\vec{\sigma}\cdot\frac{\vec{p}_{\Lambda}}{|\vec{p}_{\Lambda}|}.
\label{dllp}
\end{equation}
Then obviously, we get
\begin{equation}
[{\cal D}^{(1/2)}(\Lambda L(p))]^{-1}=(\frac{(\Lambda p)^0 +m}{2m})^{1/2}I-(\frac{(\Lambda p)^0-m}{2m})^{1/2}
\vec{\sigma}\cdot\frac{\vec{p}_{\Lambda}}{|\vec{p}_{\Lambda}|}.
\end{equation}
Here, $\vec{\sigma}=(\sigma_x, \sigma_y, \sigma_z)$ and $\sigma_i$ is the Pauli matrix.
If we put, equations (\ref{dlam}), (\ref{dlp}) and (\ref{dllp}) into eq. (\ref{dwlp}), we obtain
\begin{eqnarray}
&&{\cal D}^{(1/2)}(W(\Lambda, p))\nonumber\\
&&=\alpha^+_{\vec{p}_{\Lambda}}\alpha^+_{\vec{p}}\cosh\frac{\alpha}{2}
-\alpha^-_{\vec{p}_{\Lambda}}\alpha^+_{\vec{p}}\cosh\frac{\alpha}{2}(\vec{\sigma}\cdot\hat{p}_{\Lambda})
+\alpha^+_{\vec{p}_{\Lambda}}\alpha^+_{\vec{p}}\sinh\frac{\alpha}{2}(\vec{\sigma}\cdot\hat{e})\nonumber\\
&&-\alpha^-_{\vec{p}_{\Lambda}}\alpha^+_{\vec{p}}\sinh\frac{\alpha}{2}(\vec{\sigma}\cdot\hat{p}_{\Lambda})
(\vec{\sigma}\cdot\hat{e})+\alpha^+_{\vec{p}_{\Lambda}}\alpha^-_{\vec{p}}\cosh\frac{\alpha}{2}(\vec{\sigma}\cdot\hat{p})
-\alpha^-_{\vec{p}_{\Lambda}}\alpha^-_{\vec{p}}\cosh\frac{\alpha}{2}(\vec{\sigma}\cdot\hat{p}_{\Lambda})
(\vec{\sigma}\cdot\hat{p})\nonumber\\
&&+\alpha^+_{\vec{p}_{\Lambda}}\alpha^-_{\vec{p}}\sinh\frac{\alpha}{2}(\vec{\sigma}\cdot\hat{e})
(\vec{\sigma}\cdot\hat{p})-\alpha^-_{\vec{p}_{\Lambda}}\alpha^-_{\vec{p}}\sinh\frac{\alpha}{2}(\vec{\sigma}\cdot\hat{p}_{\Lambda})
(\vec{\sigma}\cdot\hat{e})(\vec{\sigma}\cdot\hat{p}),
\label{dwlpii}
\end{eqnarray}
where $\alpha^{\pm}_{\vec{p}}=(\frac{p^0\pm m}{2m})^{1/2}, \alpha^{\pm}_{\vec{p}_{\Lambda}}
=(\frac{(\Lambda p)^0\pm m}{2m})^{1/2}, \hat{p}=\frac{\vec{P}}{|\vec{p}|}$ and
$\hat{p}_{\Lambda}=\frac{\vec{P}_{\Lambda}}{|\vec{p}_{\Lambda}|}$.
Eq. (\ref{dwlpii}) can be rearranged into the following form:
\begin{equation}
{\cal D}^{1/2}(W(\Lambda, p))=A+B\vec{\sigma}\cdot\vec{p}+C\vec{\sigma}\cdot\hat{e}+iD\vec{\sigma}\cdot(\vec{p}\times\hat{e}),
\end{equation}
by using the relations
\begin{equation}
(\vec{\sigma}\cdot\vec{a})(\vec{\sigma}\cdot\vec{b})=\vec{a}\cdot\vec{b}+i\vec{\sigma}\cdot(\vec{a}\times\vec{b}),
\end{equation}
and
\begin{equation}
(\vec{\sigma}\cdot\hat{p}_{\Lambda})(\vec{\sigma}\cdot\hat{e})(\vec{\sigma}\cdot\hat{p})=(\hat{p}_{\Lambda}\cdot\hat{e})(\vec{\sigma}\cdot\hat{p})
+i(\hat{p}_{\Lambda}\times\hat{e})\cdot\hat{p}+\vec{\sigma}\cdot\{\hat{p}\times(\hat{p}_{\Lambda}\times\hat{e})\}.
\end{equation}
The coefficients A, B, C and D are obtained after lengthy mathematical manipulations \cite{halp}\cite{ahn}. They are
\begin{eqnarray}
A&=&\frac{1}{[(p^0 +m)((\Lambda p)^0 +m)]^{1/2}}\{(p^0 +m)\cosh\frac{\alpha}{2} + (\vec{p}\cdot\hat{e})\sinh\frac{\alpha}{2}\},\\
B&=&C=0\\
D&=&-\frac{1}{[(p^0 +m)((\Lambda p)^0 +m)]^{1/2}}\sinh\frac{\alpha}{2}.
\end{eqnarray}
Then the Wigner representation of the Lorentz group for the spin-$\frac{1}{2}$ becomes:
\begin{eqnarray}
&&{\cal D}^{(1/2)}(W(\Lambda, p))\nonumber\\
&&=\frac{1}{[(p^0 +m)((\Lambda p)^0 +m)]^{1/2}}\{(p^0 +m)\cosh\frac{\alpha}{2} +
(\vec{p}\cdot\hat{e})\sinh\frac{\alpha}{2}-i \sinh\frac{\alpha}{2}\vec{\sigma}\cdot(\vec{p}\times\hat{e})\}\\
&&=\cos\frac{\Omega_{\vec{p}}}{2}+i\sin\frac{\Omega_{\vec{p}}}{2}(\vec{\sigma}\cdot\hat{n}),
\label{rota}
\end{eqnarray}
with
\begin{equation}
\cos\frac{\Omega_{\vec{p}}}{2}=\frac{\cosh\frac{\alpha}{2}\cosh\frac{\delta}{2}+\sinh\frac{\alpha}{2}\sinh\frac{\delta}{2}(\hat{e}\cdot\hat{p})}
{[\frac{1}{2}+\frac{1}{2}\cosh\alpha\cosh\delta+\frac{1}{2}\sinh\alpha\sinh\delta(\hat{e}\cdot\hat{p})]^{1/2}},
\end{equation}
and
\begin{equation}
\sin\frac{\Omega_{\vec{p}}}{2} \hat{n}=\frac{\sinh\frac{\alpha}{2}\sinh\frac{\delta}{2}(\hat{e}\times\hat{p})}
{[\frac{1}{2}+\frac{1}{2}\cosh\alpha\cosh\delta+\frac{1}{2}\sinh\alpha\sinh\delta(\hat{e}\cdot\hat{p})]^{1/2}},
\end{equation}
where $\cosh\delta=\frac{p^0}{m}$.
We note that the eq. (\ref{rota}) indicates the Lorentz group can be represented by the pure rotation by axis $\hat{n}=\hat{e}\times\hat{p}$
for the two-component spinor.

As an example, we consider the case of momentum vector in the $z$-direction and the boost in the $x$-direction.
In this case, we have
\begin{eqnarray}
\cos\frac{\Omega_{\vec{p}}}{2}&=&\frac{\cosh\frac{\alpha}{2}\cosh\frac{\delta}{2}}{[\frac{1}{2}+\frac{1}{2}\cosh\alpha\cosh\delta]^{1/2}},\\
\sin\frac{\Omega_{\vec{p}}}{2} \hat{n}&=&\frac{-\hat{y}\sinh\frac{\alpha}{2}\sinh\frac{\delta}{2}}{[\frac{1}{2}+\frac{1}{2}\cosh\alpha\cosh\delta]^{1/2}},
\end{eqnarray}
and
\begin{eqnarray}
{\cal D}^{1/2}(W(\Lambda, p))&=&\cos\frac{\Omega_{\vec{p}}}{2}-i\sigma_y\sin\frac{\Omega_{\vec{p}}}{2}\nonumber\\
&=&\left( \begin{array}{cc}
          \cos\frac{\Omega_{\vec{p}}}{2} & -\sin\frac{\Omega_{\vec{p}}}{2}\\
          \sin\frac{\Omega_{\vec{p}}}{2} & \cos\frac{\Omega_{\vec{p}}}{2}
          \end{array} \right)
\label{wign}
\end{eqnarray}
The Wigner angle $\Omega_{\vec{p}}$ is defined by
\begin{equation}
\tan\Omega_{\vec{p}}=\frac{\sinh\alpha \sinh\delta}{\cosh\alpha+\cosh\delta}.
\label{wigan}
\end{equation}

In the Figure 1, we plot the Wigner angle given by eq. (\ref{wigan}) as function of $\beta=\frac{v}{c}$ for i)$E_{\vec{p}}/m=10$ (solid line),
ii)$E_{\vec{p}}/m=100$ (dashed line), and iii)$E_{\vec{p}}/m=1000$ (dotted line). It is interesting to note that the higher energy (at the rest frame)
of a particle for a given mass, the smaller the rotation angle at lower $\beta$, however they all reach the same limit, $\pi/2$ when $\beta$ becomes
$1$.

\section{Relativistic entanglement of quantum states and Bell's inequality.}
We define the momentum-conserved entangled Bell stats for spin-$\frac{1}{2}$ particles in the rest frame as follows:
\begin{mathletters}
\begin{eqnarray}
\Psi_{00}&=&\frac{1}{\sqrt{2}}\{
a^+(\vec{p},{{1}\over {2}})a^+(-\vec{p},\frac{1}{2})+a^+(\vec{p},-\frac{1}{2})a^+(-\vec{p},-\frac{1}{2})\}\Psi_0,\label{bello}\\
\Psi_{01}&=&\frac{1}{\sqrt{2}}\{
a^+(\vec{p},\frac{1}{2})a^+(-\vec{p},\frac{1}{2})-a^+(\vec{p},-\frac{1}{2})a^+(-\vec{p},-\frac{1}{2})\}\Psi_0,\label{bellt}\\
\Psi_{10}&=&\frac{1}{\sqrt{2}}\{
a^+(\vec{p},\frac{1}{2})a^+(-\vec{p},-\frac{1}{2})+a^+(\vec{p},-\frac{1}{2})a^+(-\vec{p},\frac{1}{2})\}\Psi_0,\label{bellth}\\
\Psi^{11}&=&\frac{1}{\sqrt{2}}\{
a^+(\vec{p},\frac{1}{2})a^+(-\vec{p},-\frac{1}{2})-a^+(\vec{p},-\frac{1}{2})a^+(-\vec{p},\frac{1}{2})\}\Psi_0,\label{bellf}\\
\nonumber
\end{eqnarray}
\end{mathletters}
where $\Psi_0$ is the Lorentz invariant vacuum state.

For an observer in another reference frame $S'$ described by an
arbitrary boost $\Lambda$, the transformed Bell states are given
by
\begin{equation}
\Psi_{ij}\rightarrow U(\Lambda)\Psi_{ij}. \label{ptrans}
\end{equation}

For example, from equations (\ref{trans}) and (\ref{bello}),
$U(\Lambda)\Psi_{00}$ becomes
\begin{eqnarray}
U(\Lambda)\Psi_{00}&=&\frac{1}{\sqrt{2}}\{
U(\Lambda)a^+(\vec{p},\frac{1}{2})U^{-1}(\Lambda)U(\Lambda)a^+(-\vec{p},\frac{1}{2})U^{-1}(\Lambda)\nonumber\\
& &+
U(\Lambda)a^+(\vec{p},-\frac{1}{2})U^{-1}(\Lambda)U(\Lambda)a^+(-\vec{p},-\frac{1}{2})U^{-1}(\Lambda)\}U(\Lambda)\Psi_0
\nonumber\\
&=&\frac{1}{\sqrt{2}}\sum_{\sigma,\sigma'}\{\sqrt{\frac{(\Lambda
p)^0}{p^0}}{\cal D}^{(\frac{1}{2})}_{\sigma\frac{1}{2}}(W(\Lambda,
p))\sqrt{\frac{(\Lambda {\cal P}p)^0}{({\cal P}p)^0}}{\cal
D}^{(\frac{1}{2})}_{\sigma'\frac{1}{2}}(W(\Lambda, {\cal
P}p))a^+(\vec{p}_{\Lambda},\sigma)a^+(-\vec{p}_{\Lambda},\sigma')\nonumber\\
&&+\sqrt{\frac{(\Lambda p)^0}{p^0}}{\cal
D}^{(\frac{1}{2})}_{\sigma-\frac{1}{2}}(W(\Lambda,
p))\sqrt{\frac{(\Lambda {\cal P}p)^0}{({\cal P}p)^0}}{\cal
D}^{(\frac{1}{2})}_{\sigma'-\frac{1}{2}}(W(\Lambda, {\cal
P}p))a^+(\vec{p}_{\Lambda},\sigma)a^+(-\vec{p}_{\Lambda},\sigma')\}\Psi_0
\label{tbell}
\end{eqnarray}
and so on.
For simplicity, we assume that $\vec{p}$ is in $z$-direction,
$\vec{p}=(0,0,p)$ and the boost $\Lambda$ is in $x$-direction.

Then from equations (\ref{wign}) and (\ref{tbell}), we obtain
\begin{mathletters}
\begin{eqnarray}
U(\Lambda)\Psi_{00}&=&\frac{(\Lambda p)^0}{p^0}\cos\Omega_{\vec{p}}
\frac{1}{\sqrt{2}}\{ a^+(\vec{p}_{\Lambda},\frac{1}{2})a^+(-\vec{p}_{\Lambda},\frac{1}{2})+a^+(\vec{p}_{\Lambda},-\frac{1}{2})a^+(-\vec{p}_{\Lambda},-\frac{1}{2})\}\Psi_0\nonumber\\
&&-\frac{(\Lambda p)^0}{p^0}\sin\Omega_{\vec{p}}\frac{1}{\sqrt{2}}\{
a^+(\vec{p}_{\Lambda},\frac{1}{2})a^+(-\vec{p}_{\Lambda},-\frac{1}{2})-a^+(\vec{p}_{\Lambda},-\frac{1}{2})a^+(-\vec{p}_{\Lambda},\frac{1}{2})\}\Psi_0\nonumber\\
&=&\frac{(\Lambda p)^0}{p^0}|\vec{p}_{\Lambda}, -\vec{p}_{\Lambda}\rangle\otimes\{\cos\Omega_{\vec{p}}\frac{1}{\sqrt{2}}(|\frac{1}{2},\frac{1}{2}\rangle+
|-\frac{1}{2}, -\frac{1}{2}\rangle)\}\nonumber\\
&&-\frac{(\Lambda p)^0}{p^0}|\vec{p}_{\Lambda}, -\vec{p}_{\Lambda}\rangle\otimes\{\sin\Omega_{\vec{p}}\frac{1}{\sqrt{2}}(|\frac{1}{2},-\frac{1}{2}\rangle-
|-\frac{1}{2}, \frac{1}{2}\rangle)\}\nonumber\\
&=&\frac{(\Lambda p)^0}{p^0}\{\cos\Omega_{\vec{p}} \Psi'_{00} -
\sin\Omega_{p} \Psi'_{11}\},\label{ulpi}
\end{eqnarray}
where $\Psi'{ij}$ is the Bell states in the moving frame $S'$
whose momenta are transformed as $\vec{p}\to\vec{p}_{\Lambda},
-\vec{p}\to-\vec{p}_{\Lambda}$. Likewise, we have
\begin{eqnarray}
U(\Lambda)\Psi_{01}&=&\frac{(\Lambda p)^0}{p^0}
\Psi'_{01},\label{ulpii}\\
U(\Lambda)\Psi_{10}&=&\frac{(\Lambda p)^0}{p^0}
\Psi'_{10},\label{ulpiii},
\end{eqnarray}
and
\begin{equation}
U(\Lambda)\Psi_{11}=\frac{(\Lambda p)^0}{p^0}\{\sin\Omega_{\vec{p}}
\Psi'_{00} +\cos\Omega_{p} \Psi'_{11}\}.\label{ulpiv}
\end{equation}
\end{mathletters}
If we regard $\Psi'_{ij}$ as Bell states
in the moving frame $S'$, then to an observer in $S'$, the effects
of the Lorentz transformation on entangled Bell
states among themselves should appear as rotations of Bell states in the
frame $S'$.
We are now, ready to check whether the Lorentz transformed Bell states violate the Bell's
inequality by calculating the average of the Bell observable defined in the equation (\ref{bell}).
Before we proceed, we note that the Bell states can be categorized into two groups. The first group is
the subset $\{\Psi_{00}, \Psi_{11}\}$ which transforms among themselves by the Lorentz transformation, and
the second group is the set $\{\Psi_{01}, \Psi_{10}\}$ which remains invariant in forms under the boost except
the momentum change. Normalized relativistic spin observables $\hat{a}, \hat{b}$ are given by \cite{czac}
\begin{equation}
\hat{a}=\frac{(\sqrt{1-\beta^2} \vec{a}_{\perp}+\vec{a}_{\parallel})\cdot\vec{\sigma}}{\sqrt{1+\beta^2[(\hat{e}\cdot\vec{a})-1]}}
\label{hata}
\end{equation}
and
\begin{equation}
\hat{b}=\frac{(\sqrt{1-\beta^2} \vec{b}_{\perp}+\vec{b}_{\parallel})\cdot\vec{\sigma}}{\sqrt{1+\beta^2[(\hat{e}\cdot\vec{b})-1]}},
\label{hatb}
\end{equation}
where the subscript $\perp$ and $\parallel$ denote the components of
$\vec{a}$ (or $\vec{b}$) which are perpendicular and parallel to the boost direction, respectively. Moreover, $|\vec{a}|=|\vec{b}|=1$.

Case I: $\Psi_{00}\rightarrow U(\Lambda)\Psi_{00}$

>From eq. (\ref{ulpi}), we have
\begin{eqnarray}
U(\Lambda)\Psi_{00}&=&\frac{(\Lambda p)^0}{p^0}|\vec{p}_{\Lambda}, -\vec{p}_{\Lambda}\rangle\otimes
[\frac{1}{\sqrt{2}}\cos\Omega_{\vec{p}}(|\frac{1}{2},\frac{1}{2}\rangle+|-\frac{1}{2}, -\frac{1}{2}\rangle)\nonumber\\
&&-\frac{1}{\sqrt{2}}\sin\Omega_{\vec{p}}(|\frac{1}{2},-\frac{1}{2}\rangle-|-\frac{1}{2}, \frac{1}{2}\rangle)].
\end{eqnarray}
Then, after some mathematical manipulations, we get
\begin{mathletters}
\begin{eqnarray}
\hat{a}\otimes\hat{b}|\frac{1}{2},\frac{1}{2}\rangle&=&\frac{1}{\sqrt{[1+\beta^2(a_x^2-1)][1+\beta^2(b_x^2-1)]}}\{(1-\beta^2)a_z b_z|\frac{1}{2},\frac{1}{2}\rangle\nonumber\\
&&+\sqrt{1-\beta^2}a_z (b_x+ib_y\sqrt{1-\beta^2})|\frac{1}{2},-\frac{1}{2}\rangle\nonumber\\
&&+\sqrt{1-\beta^2}b_z (a_x+ia_y\sqrt{1-\beta^2})|-\frac{1}{2},\frac{1}{2}\rangle\nonumber\\
&&+(a_x +i a_y\sqrt{1-\beta^2})(b_x+ib_y\sqrt{1-\beta^2})|-\frac{1}{2},-\frac{1}{2}\rangle\},\label{abtri}\\
\hat{a}\otimes\hat{b}|-\frac{1}{2},-\frac{1}{2}\rangle&=&\frac{1}{\sqrt{[1+\beta^2(a_x^2-1)][1+\beta^2(b_x^2-1)]}}\{(a_x -i a_y\sqrt{1-\beta^2})
(b_x-ib_y\sqrt{1-\beta^2})|\frac{1}{2},\frac{1}{2}\rangle\nonumber\\
&&-\sqrt{1-\beta^2}b_z (a_x-ia_y\sqrt{1-\beta^2})|\frac{1}{2},-\frac{1}{2}\rangle\nonumber\\
&&-\sqrt{1-\beta^2}a_z (b_x-ib_y\sqrt{1-\beta^2})|-\frac{1}{2},\frac{1}{2}\rangle\nonumber\\
&&+(1-\beta^2)a_z b_z|-\frac{1}{2},-\frac{1}{2}\rangle\},\label{abtrii}\\
\hat{a}\otimes\hat{b}|\frac{1}{2},-\frac{1}{2}\rangle&=&\frac{1}{\sqrt{[1+\beta^2(a_x^2-1)][1+\beta^2(b_x^2-1)]}}
\{\sqrt{1-\beta^2}a_z (b_x-ib_y\sqrt{1-\beta^2})|\frac{1}{2},\frac{1}{2}\rangle\nonumber\\
&&-(1-\beta^2)a_z b_z|\frac{1}{2},-\frac{1}{2}\rangle\nonumber\\
&&+(a_x +i a_y\sqrt{1-\beta^2})(b_x-ib_y\sqrt{1-\beta^2})|-\frac{1}{2},\frac{1}{2}\rangle\nonumber\\
&&-\sqrt{1-\beta^2}b_z (a_x+ia_y\sqrt{1-\beta^2})|-\frac{1}{2},-\frac{1}{2}\rangle\},\label{abtriii}\\
\hat{a}\otimes\hat{b}|-\frac{1}{2},\frac{1}{2}\rangle&=&\frac{1}{\sqrt{[1+\beta^2(a_x^2-1)][1+\beta^2(b_x^2-1)]}}
\{\sqrt{1-\beta^2}b_z (a_x-ia_y\sqrt{1-\beta^2})|\frac{1}{2},\frac{1}{2}\rangle\nonumber\\
&&+(a_x -i a_y\sqrt{1-\beta^2})(b_x+ib_y\sqrt{1-\beta^2})|\frac{1}{2},-\frac{1}{2}\rangle\nonumber\\
&&-(1-\beta^2)a_z b_z|-\frac{1}{2},\frac{1}{2}\rangle\nonumber\\
&&-\sqrt{1-\beta^2}a_z (b_x+ib_y\sqrt{1-\beta^2})|-\frac{1}{2},-\frac{1}{2}\rangle\}\label{abtriv}
\end{eqnarray}
\end{mathletters}
for the boost in the $x$-direction. The calculation of $\langle\hat{a}\otimes\hat{b}\rangle$ is straightforward and is given by
\begin{eqnarray}
\langle\hat{a}\otimes\hat{b}\rangle&=&\frac{1}{\sqrt{[1+\beta^2(a_x^2-1)][1+\beta^2(b_x^2-1)]}}\{[a_x b_x+(1-\beta^2)a_z b_z]\cos(2\Omega_{\vec{p}})\nonumber\\
&&-(1-\beta^2)a_y b_y -\sqrt{1-\beta^2}(a_z b_x-b_z a_x)\sin(2\Omega_{\vec{p}})\}.
\label{expe}
\end{eqnarray}

It is interesting to note that in the ultra-relativistic limit, $\beta\to 1$, equation (\ref{expe}) becomes
\begin{equation}
\langle\hat{a}\otimes\hat{b}\rangle\rightarrow \frac{a_x}{|a_x|}\cdot\frac{b_x}{|b_x|}\cos(2\Omega_{\vec{p}}),
\end{equation}
implying that the joint measurements are not correlated at all. As a result, one might suspect that the entangled state satisfies
the Bell's inequality. We now consider the vectors $\vec{a}=(\frac{1}{\sqrt{2}},-\frac{1}{\sqrt{2}},0),
\vec{a}'=(-\frac{1}{\sqrt{2}},-\frac{1}{\sqrt{2}},0), \vec{b}=(0, 1, 0), \vec{b}'=(1, 0, 0)$ which lead to the maximum
violation of the Bell's inequality in the non-relativistic domain, $\Omega_p=0$ and $\beta=0$. Then the Bell observable
for the $4$ relevant joint measurements becomes
\begin{eqnarray}
&&\langle\hat{a}\otimes\hat{b}\rangle+\langle\hat{a}\otimes\hat{b}'\rangle+\langle\hat{a}'\otimes\hat{b}\rangle-\langle\hat{a}'\otimes\hat{b}'\rangle\nonumber\\
&&=\frac{2}{\sqrt{2-\beta^2}}(\sqrt{1-\beta^2}+\cos\Omega_p).
\label{bells}
\end{eqnarray}
In the ultra-relativistic limit where $\beta=1$, the eq. (\ref{bells}) gives the maximum value of $2$ satisfying the Bell's inequality as expected.

Case II. $\Psi_{10}\rightarrow U(\Lambda)\Psi_{10}$

This case is interesting because the Lorentz transformation leaves the Bell state invariant in form, which is
\begin{equation}
U(\Lambda)\Psi_{10}=\frac{(\Lambda p)^0}{p^0}|\vec{p}_{\Lambda}, -\vec{p}_{\Lambda}\rangle\otimes\frac{1}{\sqrt{2}}(|\frac{1}{2},-\frac{1}{2}\rangle+
|-\frac{1}{2}, \frac{1}{2}\rangle).
\end{equation}
>From equations (\ref{abtri}) to (\ref{abtriv}), we obtain
\begin{equation}
\langle\hat{a}\otimes\hat{b}\rangle=\frac{1}{\sqrt{[1+\beta^2(a_x^2-1)][1+\beta^2(b_x^2-1)]}}\{a_x b_x+(1-\beta^2)(a_y b_y-a_z b_z)\}.
\label{expei}
\end{equation}
Then, in the ultra-relativistic limit, $\beta\to 1$, we have

\begin{equation}
\langle\hat{a}\otimes\hat{b}\rangle\rightarrow \frac{a_x}{|a_x|}\cdot\frac{b_x}{|b_x|},
\end{equation}
again, indicating the joint measurements, become uncorrelated in this limit. We consider the vectors
$\vec{a}=(\frac{1}{\sqrt{2}},\frac{1}{\sqrt{2}},0),
\vec{a}'=(-\frac{1}{\sqrt{2}},-\frac{1}{\sqrt{2}},0), \vec{b}=(0, 1, 0), \vec{b}'=(1, 0, 0)$ which lead to the maximum
violation of the Bell's inequality in the non-relativistic regime.
Then the Bell observable for the $4$ relevant joint measurements becomes
\begin{eqnarray}
&&\langle\hat{a}\otimes\hat{b}\rangle+\langle\hat{a}\otimes\hat{b}'\rangle+\langle\hat{a}'\otimes\hat{b}\rangle-\langle\hat{a}'\otimes\hat{b}'\rangle\nonumber\\
&&=\frac{2}{\sqrt{2-\beta^2}}(\sqrt{1-\beta^2}+1),
\label{bellsi}
\end{eqnarray}
thus giving same maximum value as in the case I. Above results show that it may be irrelevant whether the form of the entanglement is
invariant or not in calculating the Bell's inequality. It also would not be too difficult to show that one can obtain the same value for the Bell
observables given by eq. (\ref{bellsi}) for $U(\Lambda)\Psi_{01}$ and $U(\Lambda)\Psi_{11}$ implying eq. ({\ref{bellsi}) is the universal result.

In the Figure 2, the unversal Bell observable (eq. \ref{bellsi}) is calculated as a function of $\beta=\frac{v}{c}$.

We note that the Bell observable (eq. \ref{bellsi}) is
a monotonically decreasing function of $\beta$, approaching the limit value of $2$ from above when $\beta=1$ which indicates the degree
of violation of the Bell's inequality is decreasing with increasing velocity of the observer. There still remains a question of why the Bell's
inequality is satisfied at the ultra-relativistic limit. One plausible explanation seems that the
shape of the entangled pair seen by the observer becomes more deformed boost speed increases and as a consequence, both spins of
the pair are tilted toward the boost axis \cite{inpr}. Then such a spin has, for $\beta=1$, commuting components and behaves like a classical observable.

\section{summary}
In this work, we studied the Lorentz transformed entangled Bell states and the Bell observables investigate whether the Bell's
inequality is violated in all regime. We first present the quantum state transformation and the Wigner representation of Lorentz group
from the heuristic point of view. The calculated Wigner angle as a function of $\beta=\frac{v}{c}$ shows that it depends on the
energy-mass ratio. We have calculated the Bell observable for the joint $4$ measurements and found that the
results are universal for all entangled states:
\begin{eqnarray}
c(\vec{a},\vec{a}',\vec{b},\vec{b}')&=&\langle\hat{a}\otimes\vec{b}\rangle+\langle\hat{a}\otimes\vec{b}'\rangle+\langle\hat{a}'\otimes\vec{b}\rangle
-\langle\hat{a}'\otimes\vec{b}'\rangle\nonumber\\
&=&\frac{2}{\sqrt{2-\beta^2}}(1+\sqrt{1-\beta^2}),
\nonumber
\end{eqnarray}
where $\hat{a}, \hat{b}$ are the relativistic spin observables derived from the Pauli-Lubanski pseudo vector. It turn out that the Bell
observable is a monotonically decreasing function of $\beta$ and approaches the limit value of $2$ when $\beta=1$ indicating that
the Bell's inequality is satisfied in the ultra-relativistic limit.
\vspace{2.0cm}

\centerline{\bf Acknowledgements}

This work was supported by the Korean Ministry of Science and
Technology through the Creative Research Initiatives Program under
Contact No. M1-0116-00-0008. We are also indebted to M. Czachor and M. S. Kim for valuable discussions.

\appendix

\section*{ Derivation of eq.(14)}
Let $\vec{J}=(J_1, J_2, J_3)$ and $\vec{K}=(K_1, K_2, K_3)$ be generators for the rotation and the boost, respectively, and define
\begin{equation}
\vec{A}=\frac{1}{2}(\vec{J}+i\vec{K})\hspace{1.0cm}\mbox{and} \hspace{1.0cm} \vec{B}=\frac{1}{2}(\vec{J}-i\vec{K})
\label{gener}
\end{equation}
Then it is easy to show that
\begin{eqnarray}
&&[A_i, A_j]=i\epsilon_{ijk}A_k,\nonumber\\
&&[B_i, B_j]=i\epsilon_{ijk}B_k,\nonumber\\
&&[A_i, B_j]=0.
\label{comm}
\end{eqnarray}
We now define the unitary transformation corresponding to the homogeneous Lorentz transformation as
\begin{eqnarray}
U(\Lambda)&=&1+\frac{i}{2}\omega_{\mu\nu}J^{\mu\nu}\nonumber\\
&=&e^{\frac{i}{2}\omega_{\mu\nu}J^{\mu\nu}},
\label{loren}
\end{eqnarray}
with
\begin{equation}
\Lambda^{\mu}_{{}\nu}=\delta^{\mu}_{{}\nu}+\omega^{\mu}_{{}\nu}
\end{equation}
Here $\Lambda^{\mu}_{{}\nu}$ denotes the homogenous Lorentz transformation. The antisymmetric tensor $\omega_{\mu\nu}$ and the generator
$J^{\mu\nu}$ can be written as
\begin{eqnarray}
\omega_{\mu\nu}&=&\left[ \begin{array}{cccc}
                 0            & \omega_{01} & \omega_{02} & \omega_{03}\\
                 -\omega_{01} &    0        & \omega_{12} & \omega_{13}\\
                 -\omega_{02} & -\omega_{12}&     0       & \omega_{23}\\
                 -\omega_{03} & -\omega_{13}& -\omega_{23}&   0
                 \end{array}
                 \right]\nonumber\\
               &=&\left[ \begin{array}{cccc}
                 0       & \phi_1    & \phi_2   & \phi_3 \\
                 -\phi_1 &    0      & \theta_3 & -\theta_2\\
                 -\phi_2 & -\theta_3 &     0    & \theta_1\\
                 -\phi_3 & \theta_2  & -\theta_1&   0
                 \end{array}
                 \right],
\end{eqnarray}
and
\begin{eqnarray}
J^{\mu\nu}&=&\left[ \begin{array}{cccc}
                 0       & J^{01} & J^{02} & J^{03}\\
                 -J^{01} &    0   & J^{12} & J^{13}\\
                 -J^{02} & -J^{12}&     0  & J^{23}\\
                 -J^{03} & -J^{13}& -J^{23}&   0
                 \end{array}
                 \right]\nonumber\\
               &=&\left[ \begin{array}{cccc}
                 0    & K_1  & K_2 & K_3 \\
                 -K_1 &  0   & J_3 & -J_2\\
                 -K_2 & -J_3 &  0  & J_1\\
                 -K_3 & J_2  & -J_1&  0
                 \end{array}
                 \right].
\end{eqnarray}
Here,
\begin{eqnarray}
\vec{J}&=&(J_1, J_2, J_3)=(J^{23}, J^{21}, J^{12}),\nonumber\\
\vec{K}&=&(K_1, K_2, K_3)=(J^{01}, J^{12}, J^{13}),\nonumber\\
\vec{\theta}&=&(\theta_1, \theta_2, \theta_3)=(\omega_{23}, \omega_{31}, \omega_{12}),\nonumber\\
\vec{\phi}&=&(\phi_1, \phi_2, \phi_3)=(\omega_{01}, \omega_{02}, \omega_{03}).
\label{jktp}
\end{eqnarray}
Then from equations (\ref{loren}) to (\ref{jktp}), we obtain
\begin{eqnarray}
U(\Lambda)&=&\exp\frac{i}{2}\omega_{\mu\nu}J^{\mu\nu}\nonumber\\
&=&\exp[i(\omega_{01}J^{01}+\omega_{02}J^{02}+\omega_{03}J^{03}+\omega_{12}J^{12}+\omega_{23}J^{23}+\omega_{31}J^{31}]\nonumber\\
&=&\exp[i(\vec{\phi}\cdot\vec{K}+\vec{\theta}\cdot\vec{J})]\nonumber\\
&=&\exp[i(\vec{\phi}\cdot (-i)(\vec{A}-\vec{B})+\vec{\theta}\cdot(\vec{A}+\vec{B})]\nonumber\\
&=&\exp[i(\vec{\theta}-i\vec{\phi})\cdot\vec{A}+i(\vec{\theta}+i\vec{\phi})\cdot\vec{B}]\nonumber\\
&=&\exp[i(\vec{\theta}-i\vec{\phi})\cdot\vec{A}]\exp[i(\vec{\theta}+i\vec{\phi})\cdot\vec{B}],
\label{utran}
\end{eqnarray}
since $[\vec{A}, \vec{B}]=0$. From eq. (\ref{utran}), it can be seen that $U(\Lambda)$ can be represented by $SU(2)\otimes SU(2)$ for
spin-$\frac{1}{2}$ particle. From the relation $[\vec{A}, \vec{B}]=0$, we can find the common eigenstate $\psi=\psi(a_j, b_j)$ which
can be used in the representation of $U(\Lambda)$. As a special case, we consider the case of $b_j=0$ and $j=\frac{1}{2}$.
Then $\vec{B}=\frac{1}{2}(\vec{J}-i\vec{K})=0, \vec{A}=\frac{1}{2}(\vec{J}+i\vec{K})=\vec{J}$, and as result, we get
\begin{eqnarray}
U(\Lambda)&=&\exp[i(\vec{\theta}-i\vec{\phi})\cdot\vec{J}]\nonumber\\
&=&\exp[i(\vec{\theta}-i\vec{\phi})\cdot\frac{\vec{\sigma}}{2}].
\end{eqnarray}
For a given $2$-component spinor $\phi_R$, we note that $\phi_R$ transforms under the homogenous Lorentz transformation as
\begin{eqnarray}
\phi_R \rightarrow && U(\Lambda)\phi_R\nonumber\\
&&=\exp(\frac{1}{2}\vec{\sigma}\cdot\vec{\phi})\phi_R.
\end{eqnarray}
Using the relations, $\gamma=\cosh\phi$, $\gamma\beta=\sinh\phi$, and $\hat{p}=\hat{\phi}$, we finally obtain:
\begin{equation}
\phi_R(\vec{p})=[(\frac{\gamma+1}{2})^{1/2}+\vec{\sigma}\cdot\frac{\vec{p}}{|\vec{p}|}(\frac{\gamma-1}{2})^{1/2}]\phi_R(0)
\end{equation}


\end{document}